# The History of Digital Ethics

*Vincent C. Müller*

www.sophia.de

*Abstract:* Digital ethics, also known as computer ethics or information ethics, is now a lively field that draws a lot of attention. But what were the developments that led to its existence and present state? What are the traditions, concerns, and technological and social developments that guided digital ethics? How did ethical issues change with the digitalization of human life? How did the traditional discipline of philosophy respond and how was 'applied ethics' influenced by these developments? This chapter proposes to view the history of digital ethics in three phases: pre-digital modernity (before the invention of digital technology), digital modernity (with digital technology but analogue lives), and digital post-modernity (with digital technology and digital lives). For each phase, the developments in digital ethics are explained with the background of the technological and social conditions. Finally, a brief outlook is provided.

## 1. Introduction

The history of digital *ethics* as a field was strongly shaped by the development and use of digital *technologies* in society. This digital ethics often mirror the ethical concerns of the pre-digital technologies that were replaced, but in more recent times, digital technologies have also posed questions that are truly new. When 'data processing' became a more common activity in industry and public administration in the 1960s, the concerns of ethicists were old issues like *privacy, data security*, and *power* through information access. Today, digital ethics involves old issues that took on a new quality due to digital technology, such as *surveillance, news*, or *dating*, but it also covers new issues that did not exist at all, such as *automated weapons*, *search engines*, *automated decision-making*, and *existential risk from artificial intelligence (AI)*.

The terms used to name the expanding discipline have also changed over time: we started with 'computer ethics' (Bynum 2001; Johnson 1985; Vacura 2015), then more abstract terms like 'information ethics' were proposed (Floridi 1999), and now some use the term 'digital ethics' (Capurro 2010), as this Handbook does. We also have digital ethics for particular areas, such as 'the ethics of AI', 'data ethics', 'robot ethics', etc.

There are reasons for these changes: 'computer ethics' now sounds dated because it focuses attention on the machines, which made good sense when they were visible, big boxes but began to make less sense when many technical devices invisibly included computing and the location of the processor became irrelevant. The more ambitious notion of 'information ethics' involves a digital ontology (Capurro 2006) and faces a significant challenge to explain the role of the notion of 'information'; see (Floridi 1999) versus (Floridi and Taddeo 2016). Also, the term 'information ethics' is sometimes used in contexts in which information is not computed, for example, in 'library and information science'. Occasionally, one hears the term 'cyberethics' (Spinello 2020), which specifically deals with the connected 'cyberspace'—probably now an outdated term, at least outside the military. In this confusion, some people use 'digital' as the new term, which captures the most relevant phenomena and moves away from the machinery to their use. One might argue that the process of 'computing' is still fundamental but that we will probably soon care less about whether a device uses computing (analogue or digital)—like we do not care much which energy source the engine in a car uses. The notion of 'data' will continue to make sense, but, in the future, I suspect that terms like 'computing' and 'digital' will just merge into 'technology'.

Given that this Handbook already has articles on the current state of the art, this article tries to provide historical context, both in debates during the early days of information technology (IT) from the 1940s to the 1970s, when IT was an expensive technology available only in well-funded central 'computation centres'; then roughly the 1980s to the early 2000s, with networked personal computers entering offices and households; finally, the past fifteen years or so with 'smart' phones and other 'smart' devices being used privately—for new purposes that emerge with the devices.

This article is structured around two ideas, namely, that (a) technology drives ethics and (b) many issues that are now part of 'digital ethics' predate digital technology. There is a certain tension between these two ideas, however, so the discussion will try to disentangle when and in what sense 'technology drives ethics' (e.g. by posing new problems, by revealing old ones, or even by effecting ethical change) and when that 'drive' is specific to 'digital' (computing) technology. I start on the assumption that (b) is true, thus the article must begin before the invention of digital technology, in fact, even before the invention of writing. We will return to these two ideas in the conclusion.

I propose to divide history into three main sections: pre-digital modernity (before the invention of digital technology), digital modernity (with digital technology but analogue lives), and digital post-modernity (with digital technology and digital lives). The hope is that this organization matches the social developments of these periods, but I make no claim that the terminology used here is congruent with a standard history of digital society. In each section, we will briefly look at the technology and then at digital ethics. Finally, it may be mentioned that there are significant research desiderata in the field; a detailed history of digital ethics, and indeed of applied or practical ethics, is yet to be written.

## 2. Pre-digital modernity: Talking and writing

### 2.1. Technology and society

A fair proportion of the concerns of classical digital ethics are about informational privacy, information security, power through information, etc. These issues existed long before the computing age, in fact before writing was invented—after all, they also feature in village gossip.

One significant step in this timeline, however, was the beginning of symbols and iconic representations from cave paintings onwards (cf. Sassoon and Gaur 1997). These allowed records that do not immediately vanish to be maintained, as speech does, some of which can be transported to another place. It may be useful to differentiate (a) representation *for* someone, or *intentional representation*, and (b) *representation per se*, when something represents something else because that is its function in a system (assuming this is possible without intentional states). The word 'tree', pronounced by someone, is an intentional representation (type 1); the non-linguistic representation of a tree in the brain of an organism that sees the tree is a non-intentional representation (type 2) (Müller 2007). Evidently, one major step that is relevant for digital ethics was the invention and use of *writing*—for the representation of natural language but also for mathematics and other purposes. Symbols in writing are already digital; that is, they have a sharp boundary with no intermediate stages (something is either an 'A' or a 'B', it cannot be a bit of both) and they are perfectly reproducible—one can write the exact same word or sentence more than once.

In a further step, the replication of writing and images in print multiplies the impact that goes with that writing—what is printed can be transported, remembered, and read by many people. It can become more easily part of the cultural heritage. A further major step is the transmission of speech and symbols over large distances and then to larger audiences through telegraph, mail, radio, and TV. Suddenly, a single person speaking could be heard and even seen by millions of others around the globe, even in real time.

### 2.2. Ethics

There is a significant body of ethical and legal discussion on pre-digital information handling, especially after the invention of writing, printing, and mass communication. Much of it is still the law today, such as the privacy of letters and other written communication, the press laws, and laws on libel (defamation). The privacy of letters was legally protected in the early days of postal services in the early eighteenth century, for example, in the 'Prussian New Postal Order' of 1712 (Matthias 1812: 54). Remarkably, several of these laws have lost their teeth in the digital era without explicit legal change. For example, email is often not protected by the privacy of letters, and online publications are often not covered by press law.

The central issue of privacy, often connected with 'data protection', started around 1900 (Warren and Brandeis 1890), developed into a field (Hoffman 1973; Martin 1973; Westin 1968) and is still a central topic of discussion today; from classical surveillance (Macnish 2017), governance (Bennett and Raab 2003), and ethical analysis (Roessler 2017; van den Hoven et al. 2020) to analysis for activism (Véliz 2020). This is an area where the law has not caught up with technical developments in such a way that the original intentions could be maintained—it is not even clear that these intentions are still politically desired.

The power of information and misinformation was well understood after the invention of printing but especially after the invention of mass media like radio and TV and their use in propaganda—media studies and media ethics became standard academic fields after the Second World War. Media ethics is still an important aspect of digital ethics (Ess 2014), especially the aspect of the 'public sphere' (Habermas 1962).

Apart from this tradition of more 'societal' ethics, there is a more 'personal' kind of ethics of *professional responsibility* that started in this area—and had an impact in the digital era. The influential *Institute of Electrical and Electronics Engineers* (IEEE, initially American Institute of Electrical Engineers, AIEE) adopted its first 'Principles of Professional Conduct for the Guidance of the Electrical Engineer' in 1912 (AIEE 1912). 'Engineering ethics' is thus older than ethics of computing—but, interestingly, the electrical and telephone industries in the United States managed to get an exception to the demand that engineers hold a professional licence (PE). This move may have had a far-reaching impact into the computer science of today, which usually does not see itself as a discipline of engineering, and bound by the ethos of engineers—though there are computer scientists that would want to achieve recognition as a profession and thus the ethos of 'being a good engineer' (in many countries, engineering has high status and computer science degrees are 'diplomas in engineering').

Up to this point, we see the main ethical themes of privacy and data security, power of information, and professional responsibility.

### 2.3. Digital modernity: Digital ethics in IT

#### 2.3.1. Technology and society

As a rough starting point in this part of the timeline, one should take the first design for a universal computer with Babbage's 'analytic engine' in about 1840; the first actual universal computer was feasible only when computers could use electronic parts, starting with Zuse's Z3 in 1941, followed by the independently developed ENIAC in 1945, and the Manchester Mark I in 1949 and then many more machines, mostly due to military funding (Ifrah 1981). All major computers since then have been electronic universal digital computers with stored programs. Shortly after the Second World War came the beginnings of the science of 'informatics' with 'cybernetics' (Ashby 1956; Wiener 1948) and C.E. Shannon's 'A Mathematical Theory of Communication' (Shannon 1948). In 1956, J. McCarthy, M.L. Minsky, N. Rochester, and C.E. Shannon organized the Dartmouth conference on 'Artificial Intelligence', thus coining the term (McCarthy et al. 1955). Less than ten years later, H. Simon predicted, 'Machines will be capable, within 20 years, of doing any work that a man can do' (Simon 1965: 96). In 1971, integrated processor (microprocessor) computers started, with all integrated circuits in one microchip. This technology effectively started the modern computer era. Up to that point, computers had been big and very expensive devices, only used by large corporations, research centres, or public entities for 'data processing'; from the 1980s, 'personal computers' were possible (and had to be labelled as such).

Ray Kurzweil has put the development from the Second World War to the present with characteristic panache:

> Computers started out as large remote machines in air-conditioned rooms tended by white coated technicians. Subsequently they moved onto our desks, then under our arms, and now in our pockets. Soon, we'll routinely put them inside our bodies and brains. Ultimately we will become more nonbiological than biological. (Kurzweil 2002)

*2.3.2. Ethics*

### A.  Professional ethics

The first discussions about ethics and computers in digital modernity were about the personal ethics of the people who work professionally in computing—what they should or should not do. In that phase, a computer scientist was an expert, rather like a doctor or a mechanical engineer, and the question arose whether the new 'profession' needed ethics. These early discussions of computer ethics often had a certain tinge of moralizing, of having discovered an area of life that had escaped the attention of ethicists so far, but where immorality, or at least some impact on society, looms. In contrast to this, professional ethics today often take the more positive approach that practitioners face ethical problems that expert analysis might help to resolve. This suspicion of immorality was often supported by the view of practitioners that our technology is neutral and our aims laudable, thus 'ethics' is not needed—a naïve view one finds even today.

The early attempts at professional ethics moved into computer science quite early in the discipline; for example, the US *Association for Computing Machinery* (ACM) adopted 'Guidelines for Professional Conduct in Information Processing; in 1966 and Donn Parker pushed this agenda in his discipline in the ensuing years (Parker 1968). The current version is called the 'ACM Code of Ethics and Professional Conduct' (ACM 2018).

### B.  Responsible technology

The use of nuclear (atomic) bombs in the Second World War and the discussion about the risk of generating electricity in nuclear power stations from the late 1950s fuelled the increasing concern about the limits of technology in the 1960s. This political development is closely connected to the political developments in 'the generation of 1968' on the political left in Europe and the United States. The 'Club of Rome' was and is a group of high-level politicians, scientists, and industry leaders that deals with the basic, long-term problems of humankind. In 1972, it published the highly influential book, *The Limits to Growth: A Report for the Club of Rome's Project on the Predicament of Mankind* (Club of Rome 1972). It argued that the industrialized world was on an unsustainable trajectory of economic growth, using up finite resources (e.g. oil, minerals, farmable land) and increasing pollution, with the background of an increasing world population. These were the views of a radical minority at the time, and even today they are still far from commonplace.

This report and other similar discussions fuelled a generally more critical view of technology and the growth it enables. They led to a field of 'technology assessment' in terms of long-term impacts that has also dealt with information technologies (Grunwald 2002). This area of the social sciences is influential in political consulting and has several academic institutes (e.g. the Karlsruhe Institute of Technology). At the same time, a more political angle of technology is taken in the field of 'Science and Technology Studies' (STS), which is now a sizable academic field with degree programmes, journals, and conferences. As books like *The Ethics of Invention* (Jasanoff 2016) show, concerns in STS are often quite similar to those in ethics, though typically with a more 'critical' and more empirical approach. Despite these agreements, STS approaches have remained oddly separate from the ethics of computing.

Concerns about *sustainable development*, especially with respect to the environment, have been prominent on the political agenda for about forty years and they are now a central policy aim in most countries, at least officially. In 2015, the United Nations adopted the '2030 Agenda for Sustainable Development' (United Nations 2015) with seventeen 'Sustainable Development Goals'. These goals are now quite influential; for example, they guide the current development of official European Union policy on AI. The seventeen goals are: (1) no poverty; (2) zero hunger; (3) good health and well-being; (4) quality education; (5) gender equality; (6) clean water and sanitation; (7) affordable and clean energy; (8) decent work and economic growth; (9) industry, innovation, and infrastructure; (10) reducing inequality; (11) sustainable cities and communities; (12) responsible consumption and production; (13) climate action; (14) life below water; (15) life on land; (16) peace, justice, and strong institutions, and (17) partnerships for the Goals.

### C. Control

It had also been understood by some that science and engineering generally pose ethical problems. The prominent physicist, C.F. v. Weizsäcker predicted in 1968 that computer technology will fundamentally transform our lives in the coming decades (Weizsäcker 1968). Weizsäcker asked how we can have individual freedom in such a world, 'i.e. freedom from the control of anonymous powers' (439). At the end of his article, he demands a Hippocratic oath for scientists. Soon after, Weizsäcker became the founding Director of the famous *Max Planck Institute for Research into the Life in a Scientific-Technical World*, co-directed by Jürgen Habermas since 1971. At that time, there was clearly a sense with major state funders that these issues deserved their own research institute.

In the United States, the ACM had a Special Interest Group 'Computers & Society' (SIGCAS) from 1969—it is still a significant actor today and still publishes the journal *Computers and Society*. Norbert Wiener had warned of AI even before the term was coined (see Bynum 2008: 26–30; 2015). In *Cybernetics*, Wiener wrote:

> [...] we are already in a position to construct artificial machines of almost any degree of elaborateness of performance. Long before Nagasaki and the public awareness of the atomic bomb, it had occurred to me that we were here in the presence of another social potentiality of unheard-of importance for good and for evil. (Wiener 1948: 28)

Note that the atomic bomb was a starting point for a critical view on technology in his case, too. In his later book, *The Human Use of Human Beings*, he warns of manipulation:

> [...] such machines, though helpless by themselves, may be used by a human being or a block of human beings to increase their control over the rest of the race or that political leaders may attempt to control their populations by means not of machines themselves but through political techniques as narrow and indifferent to human possibility as if they had, in fact, been conceived mechanically. (Wiener 1950)

Thus, in this phase, professional responsibility gains prominence as an issue, the notion of *control* through information and machinery comes up as a theme, and there is a general concern about the longer-term impacts of technology.

## 3. Post-modernity

### 3.1. Technology and society

In this part of the timeline, from 1980 to today (2021), I will use a typical university student in a wealthy European country as an illustration. I think this timeline is useful because it is easy to forget how the availability and use of computers have changed in the past decades and even the past few years. (If this text is read a few years after writing, it will seem quaintly old-fashioned.) We will see that this is the phase in which computers enter peoples' lives and digital ethics becomes a discipline.

In the first half of the 1980s, a student would have seen a 'personal computer' (PC) in a business context, and towards the end of the 1980s they would probably own one. These PCs were not connected to a network, unless on university premises, so data exchange was through floppy disks. Floppy disks held 360KB, later 720 KB and 1.44 MB; if the PC had a hard drive at all, it would hold ca. 20–120 MB. After 1990, if private PCs had network connections, that would be through modem dial-in on analogue telephone lines that would mainly serve links to others in the same network (e.g. CompuServe or AOL), allowing email and file-transfer protocol (ftp).

Around the same time, personal computers moved from a command-line to a graphic interface, first on MacOS, then on MS Windows and UNIX. Students would use electrical typewriters or university-owned computers for their writing until *ca.* the year 2000, and often even later. The first worldwide web (WWW) page came online in 1990 and institutional web pages became common in the late 1990s; around the same time a dial-in internet connection at home through a modem became affordable, and Google was founded (1998). After 2000, it became common for a student to have a computer at home with an internet connection, though file exchanges would still be mostly via physical data carriers. By *ca.* 2010, the internet connection would be 'always on' and fast enough for frequent use of www pages, and video; by *ca.* 2019, it would be fully digital (ISDN, ASDL, . . .) and its files would often be stored in the 'cloud', that is, spaces somewhere on the internet. Fibre-optic lines started to be used around 2020. With the COVID-19 pandemic over 2020–2022, cooperative work online through live video became common.

Mobile phones (cell phones) became commonly affordable by students in the late 1990s, but these were just phones, increasingly miniaturized. The first 'smart' phone, the iPhone, was introduced in 2007. Around 2015, a typical student would own such a smartphone and would use that phone mostly for things other than calls; essentially as a portable tablet computer with wi-fi capability (but it would be called a 'phone', not a 'computer'). After 2015, the typical smartphone would be connected to the internet at all times (with 3G). The frequent use of the web-over-phone internet became affordable around 2018/2019 (with 4G), so around 2020 video calls and online teaching became possible and useful.

The students born after *ca.* 1980 (i.e. at university from around 2020) are often called 'digital natives', meaning that their teenage and adult lives took place when digital information processing was commonplace. To digital natives, pre-digital technologies like print, radio, or television, feel 'old', while for the previous generations, digital technologies feel 'new'. This generational difference may also be one of the few cases where technological change drives actual ethical change, for example, in that digital natives are not worried about privacy in the way older generations are.

Together with smartphones, we now (2022) also begin to have other 'smart' devices that incorporate computers and are connected to the internet (soon with 5G), especially portables, TVs, cars, and homes—also known as the 'Internet of Things' (IoT). 'Smart' superstructures like grids, cities, and roads are being deployed. Sensors with digital output are becoming ubiquitous. In addition, a large part of our lives is digital (and thus does not need to be captured by sensors), much of it conducted through commercial platforms and 'social media' systems. All these developments enable a surveillance economy where data is a valuable commodity (as discussed in other chapters in this Handbook).

While a 'computer' was easily recognized as a physical box until *ca.* 2010, it is now incorporated into a host of devices and systems and often not perceived as such; perhaps even designed not to be noticed (e.g. in order to collect data). Much of computing has become a transparent technology in our daily lives: we use it without special learning and do not notice its existence or that computing takes place: 'The most profound technologies are those that disappear' (Weiser 1991: 94).

For the purposes of digital ethics, the crucial developments of our students were the move from computers 'somewhere else' to their own PC (*ca.* 1990), the use of the WWW (*ca.* 1995) and their smartphone (*ca.* 2015); the current development is the move to computing as a 'transparent technology'.

### 3.2. Ethics

#### 3.2.1. Establishment

The first phase of digital ethics, or computer ethics, was the effort in the 1980s and 1990s to establish that there *is* such a thing or that there *should be* such a thing—both within philosophy or applied ethics and within computer science, especially the curriculum of computer science at universities. This 'establishment' is of significant importance for the academic field since, once 'ethics' is an established component of degrees in computer science and related disciplines, there is a labour market for academic teachers, a demand for writing textbooks and articles, etc. (Bynum 2010). It is not an accident that the field was established beyond 'professional ethics' and general societal concerns around the same time as the move of computers from labs to offices and homes occurred.

The first use of 'computer ethics' was probably by Deborah Johnson in her paper 'Computer Ethics: New Study Area for Engineering Science Students', where she remarked, 'Computer professionals are beginning to look toward codes of ethics and legislation to control the use of software' (Johnson 1978). Sometimes (Bynum 2001), it is Walter Maner who is credited with the first use for 'ethical problems aggravated, transformed or created by computer technology' (Maner 1980). Again, professional ethics seems to have been the forerunner for computer ethics, generally.

A few years later, with fundamental publications like James H. [Jim] Moor's 'What is Computer Ethics?' (Moor 1985), the first textbook (Johnson 1985), and three anthologies with established publishers (Blackwell, MIT Press, Columbia University Press), one can speak of an established small discipline (Moor and Bynum 2002). The two texts by Moor and Johnson are still the most cited works in the discipline, together with classic texts on privacy, such as (Warren and Brandeis 1890) and (Westin 1968). As (Tavani 1999) shows, in the next fifteen years there was a steady flow of monographs, textbooks, and anthologies. In the 1990s, 'ethics' started to gain a place in many computer science curricula.

In terms of *themes*, we have the classical ones (privacy, information power, professional ethics, impact of technology) and we now have increasing confidence that there is 'something unique' here. Maner says, 'I have tried to show that there are issues and problems that are unique to computer ethics. For all of these issues, there was an essential involvement of computing technology. Except for this technology, these issues would not have arisen, or would not have arisen in their highly altered form' (Maner 1996).

We now get a wider notion of digital ethics that includes issues which *only* come up in ethics of *robotics and AI*, for example, manipulation, automated decision-making, transparency, bias, autonomous systems, existential risk, etc. (Müller 2020). The relationship between robots or AI systems and humans had already been discussed in Putnam's classic paper 'Robots: Machines or Artificially Created Life?' (Putnam 1964) and it has seen a revival in the discussion of singularity (Kurzweil 1999) and existential risk from AI (Bostrom 2014).

Digital ethics now covers the human *digital life*, online and with computing devices—both on an individual level and as a society, for example, social networks (Vallor 2016). As a result, this handbook includes themes like human–robot interaction, online interaction, fake news, online relationships, advisory systems, transparency and explainability, discrimination, nudging, cybersecurity, and existential risk—in other words, the digital life is prominently discussed here; something that would not have happened even five years ago.

### 3.2.2. Institutional

The journal *Metaphilosophy*, founded by T.W. Bynum and R. Reese in 1970, first published articles on computer ethics in the mid-1980s. The journal *Minds and Machines,* founded by James Fetzer in 1991, started publishing ethics papers under the editorship of James H. Moor (2001–2010). The conference series ETHICOMP (1995) and CEPE (1997) started in Europe, and specialized journals were established: the *Journal of Information Ethics* (1992), *Science and Engineering Ethics* (1995), *Ethics and Information Technology* (1999), and *Philosophy & Technology* (2010). The conferences on 'Computing and Philosophy' (CAP), since 1986 in North America, later in Europe and Asia, united to the 'International Association for Computing and Philosophy' (IACAP) in 2011 and increasingly have a strong division on ethical issues; as do the 'Society for the Study of Artificial

Intelligence and the Simulation of Behaviour' (AISB) (in the UK) and the 'Philosophy and Theory of Artificial Intelligence' (PT-AI).

Within the academic field of philosophy, applied ethics and digital ethics have remained firmly marginal or specialist even now, with very few presentations at mainstream conferences, publications in mainstream journals, or posts in mainstream departments. As far as I can tell, no paper on digital ethics has appeared in places like the *Journal of Philosophy, Mind, Philosophical Review, Philosophy & Public Affairs* or *Ethics* to this day—while, significantly, there are papers on this topic in *Science, Nature*, or *Artificial Intelligence*. Practically orientated fields in philosophy are treated largely as the poor and slightly embarrassing cousin who has to work for a living rather than having old money in the bank. In traditional philosophy, what counts as 'a problem' is still mostly defined through tradition rather than permitting a problem to enter philosophy from the outside. Cementing this situation, few of these 'practical' fields have the ambition to have a real influence on traditional philosophy; but this is changing, and I would venture that this influence will be strong in the decades to come. It is interesting to note that the citation counts of academics in computing ethics and theory have surpassed those of comparable philosophers in related traditional areas, and similar trends are happening now with journals. One data point: as of 2020, the average article in *Mind* is cited twice within four years, while the average article in *Minds and Machines* is cited three times within four years—the number for the latter journal doubled in three years.

Several prominent philosophers have worked on theoretical issues around AI and computing (e.g. Dennett, Dreyfus, Fodor, Haugeland, Searle), typically with a foundation of their careers in related areas of philosophy, such as philosophy of mind, philosophy of language, or logic. This also applies to Jim Moor, who was one of the first people in digital ethics to hold a professorship at a reputed general university (Dartmouth College). Still, the specialized researchers in the field were at marginal institutions or doing digital ethics on the side. This changed slowly; for example, several technical universities had professors working in digital ethics relatively early on; the Technical Universities in the Netherlands founded a 4TU Centre for Ethics and Technology in 2007 (Delft, Eindhoven, Twente, and Wageningen). In the past decade, Floridi and Bostrom were appointed to professorships at Oxford, at the Oxford Internet Institute (OII) and the Future of Humanity Institute (FHI). Coeckelbergh was appointed to a chair at the philosophy department in Vienna in 2015 (where Hrachovec was already active). A few more people were and are active in philosophical issues of 'new media', for example, Ch. Ess, who moved to Oslo in 2012. The ethics of AI became a field only quite recently, with the first conference in 2012 (Artificial General Intelligence (AGI)-Impacts), but it now has its own institutes at many mainstream universities.

In other words, only five years ago, almost all scholars in digital ethics were at institutions marginal to mainstream philosophy. It is only in those last couple of years that digital ethics is becoming mainstream; many more jobs are advertised, senior positions are available to people in the field, younger faculties are picking up on the topic, and more established faculties at established institutions are beginning to deem these matters worthy of their attention. That development is rapidly gaining pace now.

I expect that mainstream philosophy will quickly pick up digital ethics in the coming years—the subject has shown itself to be mature and fruitful for classical philosophical issues, and there is an obvious societal demand and significant funding opportunities. Probably there is also some hype already. In the classic notion of a 'hype cycle' for the expectations from a new technology, the development is supposed to go through several phases: After its beginnings at the 'technology trigger', it gains more and more attention, reaching a 'peak of inflated expectations', after which a more critical evaluation begins and the expectations go down, eventually reaching a 'trough of disillusionment'. From there, a realistic evaluation shows that there is some use, so we get the 'slope of enlightenment' and eventually the technology settles on a 'plateau of productivity' and becomes mainstream. The *Gartner Hype Cycle for AI, 2019* (Goasduff 2019) sees digital ethics itself at the 'peak of inflated expectations' . . . meaning that it is downhill from here, for some time, until we hopefully reach the 'plateau of productivity'. (My own view is that this is wrong since we are seeing the beginnings of AI policy and stronger digital ethics now.)

## 4. Future

The state of the art at the present and an outlook into the future are given in the chapters of this Handbook. Moor saw a bright future even twenty years ago: 'The future of computer ethics: You ain't seen nothin' yet!' (Moor 2001), and he followed up with a programmatic plea for 'machine ethics' (Moor 2006). Moor opens the former article with the bold statement:

> Computer ethics is a growth area. My prediction is that ethical problems generated by computers and information technology in general will abound for the foreseeable future. Moreover, we will continue to regard these issues as problems of computer ethics even though the ubiquitous computing devices themselves may tend to disappear into our clothing, our walls, our vehicles, our appliances, and ourselves. (Moor 2001: 89)

The prediction has undoubtedly held up until now. The ethics of the design and use of computers is clearly an area of very high societal importance and we would do well to catch problems early on—this is something we failed to do in the area of privacy (Véliz 2020) and some hope that we will do in the area of AI (Müller 2020).

However, as Moor mentions, there is also a very different possible line that was developed around the same time: Bynum reports on an unpublished talk by Deborah G. Johnson with the title 'Computer Ethics in the 21st Century' at the 1999 ETHICOMP conference:

> On Johnson's view, as information technology becomes very commonplace—as it gets integrated and absorbed into our everyday surroundings and is perceived simply as an aspect of ordinary life—we may no longer notice its presence. At that point, we would no longer need a term like 'computer ethics' to single out a subset of ethical issues arising from the use of information technology. Computer technology would be absorbed into the fabric of life, and computer ethics would thus be effectively absorbed into ordinary ethics. (Bynum 2001: 111ff) (cf. Johnson 2004)

On Johnson's view, we will have applied ethics and the ethics will concern most themes, such as 'information privacy' or 'how to behave in a romantic relationship' (Nyholm et al. 2022)—and much of this will be taking places with or through computing devices, but it will not matter (even though many things will remain that cannot be done without such devices). In other words, the 'drive' of technology we have seen in this history will come to a close, and the technology will become transparent. This transparency will likely have ethical problems itself—it enables surveillance and manipulation. If Johnson is right, however, we will soon have the situation that all too much is digital and transparent, and thus digital ethics is in danger of disappearing into general applied ethics. In Molière's play, this bourgeois who wants to become a gentleman tells his 'philosophy master':

> Oh dear! For more than forty years I have been speaking prose while knowing nothing of it, and I am most obliged to you for telling me so. - Molière, *Le Bourgeois gentilhomme* (Act II) 1670

## 5.  Conclusion and questions

One feature that is characteristic of the new developments in digital ethics and in applied philosophy generally is how a problem becomes a problem worth investigating. In traditional philosophy, the criterion is often that there already exists a discussion in the past noting that there is something philosophically *interesting* about it, something unresolved. Thus, typically, we do not need to ask again whether that problem is worth discussing or whether it relies on assumptions we should not make (so we will find people who seriously ask whether Leibniz or Locke was right on the origin of ideas, for example). In digital ethics, what counts as a problem also includes the demand to be philosophically *interesting*, but more importantly, whether it has *relevance*. Quite often, this means that the problem first surfaces in fields other than philosophy. The initially dominant approach of *professional ethics* had a touch of 'policing' about it, of checking that everyone behaves—that moralizing gives ethics a bad name and it typically comes too late. More modern digital ethics tries to make people sensitive in the design process ('ethics by design') and to pick up problems where people really do not know what the ethically right thing to do is—these are the proper ethical problems that deserve our attention.

For the relation between ethics and computer ethics, Moor seemed right in this prediction:

> The development of ethical theory in computer ethics is sometimes overstated and sometimes understated. The overstatement suggests that computer ethics will produce a new ethical theory quite apart from traditional ethical notions. The understatement suggests that computer ethics will disappear into ordinary ethics. The truth, I predict, will be found in the middle [. . .] My prediction is that ethical theory in the future will be recognizable but reconfigured because of work done in computer ethics during the coming century. (Moor 2001: 91)

In my view, philosophers must do more than *export* an expertise from philosophy or ethics to practical problems: we must also *import* insights from these debates back to philosophy. The field of digital ethics can feed largely on societal demand and the real impact philosophical insights can have in this area, but in order to secure its place within philosophy, we must show that the work is both technically serious and has real potential to shed light on traditional issues. As an example, consider the question of when an artificial agent truly *is* an agent that is responsible for their actions—that discussion seems to provide a new angle to the debates on agency that traditionally focused on human beings. We can now ask the conceptual question anew and provide evidence from experiments with *making* things, rather than from passive observation.

Nearly 250 years ago, Immanuel Kant stated that the four main questions of philosophy are: '1. What can I know? 2. What should I do? 3. What can I hope for? 4. What is the human?' (Kant 1956/1800: 26) (questions 1–3 in Kant 1956/1781:

A805 and B33). The philosophical reflection on digital technology contributes to all four of these.

## 6. Acknowledgements

I am grateful to Karsten Weber and Eleftheria Deltsou for useful comments and to Carissa Véliz, Guido Löhr, Maximilian Karge, and Jeff White for detailed reviewing.